\documentclass{PoS}

\usepackage{url} 

\newcommand{\ecap}{	Erlangen Centre for Astroparticle Physics (ECAP),
					Friedrich-Alexander-Universit\"{a}t Erlangen-N\"{u}rnberg,
					Erwin-Rommel-Stra{\ss}e 1, 91058 Erlangen, Germany}

\title{Super-ORCA: Measuring the leptonic CP-phase with Atmospheric Neutrinos and Beam Neutrinos}

\ShortTitle{Super-ORCA}

\author{\speaker{Jannik~Hofest\"adt}, Marc Bruchner, and Thomas Eberl\\
       \ecap\\
       E-mails: 	\email{jannik.hofestaedt@fau.de}, 
               	\email{thomas.eberl@fau.de}}

\abstract{
Studying the atmospheric neutrino oscillation probabilities below 2 GeV with a multi-megaton Cherenkov detector allows for a measurement of the leptonic CP-phase $\delta_{CP}$. The most relevant CP-sensitive energy range is below the neutrino detection threshold of KM3NeT/ORCA, which is an underwater Cherenkov detector optimised to determine the neutrino mass ordering by measuring the oscillation pattern of 3-30 GeV atmospheric neutrinos. With Super-ORCA, a $\sim 10$ times more-densely instrumented version of ORCA, the detection threshold can be lowered and the event reconstruction capabilities improved.

In this paper, the key detector performance indicators for a possible Super-ORCA detector and the sensitivity to $\delta_{CP}$ with atmospheric neutrinos are presented.
Including systematics, a 1$\sigma$-resolution on $\delta_{CP}$ of about $38^\circ$ ($23^\circ$) is achieved for $\delta_{CP}=0$ ($\delta_{CP}=\pi/2$) after 10 years.
In addition, the potential of using a neutrino beam from the Protvino accelerator facility to the Super-ORCA detector is discussed.
With this, a 1$\sigma$-resolution on $\delta_{CP}$ of about $10^\circ$ ($16^\circ$) is achieved for $\delta_{CP}=0$ ($\delta_{CP}=\pi/2$) after 10 years.
}

\FullConference{36th International Cosmic Ray Conference -ICRC2019-\\
		July 24th - August 1st, 2019\\
		Madison, WI, U.S.A.}

\catcode`\@=11 
\def\parenbar{\mathpalette\p@renb@r}
\def\p@renb@r#1#2{\vbox{%
		\ifx#1\scriptscriptstyle \dimen@.7em\dimen@ii.2em\else
		\ifx#1\scriptstyle \dimen@.8em\dimen@ii.25em\else
		\dimen@1em\dimen@ii.4em\fi\fi \offinterlineskip
		\ialign{\hfill##\hfill\cr
			\vbox{\hrule width\dimen@ii}\cr
			\noalign{\vskip-.3ex}%
			\hbox to\dimen@{$\mathchar300\hfil\mathchar301$}\cr
			\noalign{\vskip-.3ex}%
			$#1#2$\cr}}}
\catcode`\@=12 
\def\nuan{\parenbar{\nu}\kern-0.4ex}

\begin{document}

\section{Introduction}
All neutrino oscillation parameters of the $3\nu$ framework are by now 
measured to a fair precision, expept for the neutrino mass ordering (NMO) and the Dirac CP-phase $\delta_{CP}$. 
The latter is associated to a possible violation of the charge-parity (CP) symmetry in neutrino flavour mixing. 
Discovering CP violation in the lepton sector and subsequently measure the Dirac CP-phase are among the main objectives in particle physics.
Recent analyses \cite{global_fit} of global data favour normal hierarchy (NO) over inverted hierarchy (IO) and $\delta_{CP} \approx 3/2\pi$, i.e. maximal CP violation, however with small significances, so that most of the $\delta_{CP}$ range is still allowed at the $3\,\sigma$-level.

In general, the favoured way to determine $\delta_{CP}$ is by doing next-generation long-baseline experiments featuring a neutrino beam facility
together with a near detector and a far detector to measure the neutrino beam properties after oscillations.
The main players among the proposed experiments are: T2HK \cite{T2HK_2015} and DUNE \cite{DUNE}.

Atmospheric neutrinos offer an alternative possibility to measure $\delta_{CP}$. 
Information on $\delta_{CP}$ is encoded in the oscillation probabilities of up-going atmospheric neutrinos that have traversed the Earth.
Theoretical and phenomenological aspects of CP violation in atmospheric neutrinos have been explored in \cite{RazzaqueSmirnov_CP} and references therein. 
The most relevant CP-sensitive energy range is $\lesssim 2$\,GeV.
As CP-violation effects are small for atmospheric neutrinos, 
a very large volume neutrino detector with sufficiently low neutrino energy detection threshold and good event reconstruction capabilities, in particular for $\nu_e$ vs $\nu_\mu$ separation, is required.

The multi-megaton underwater Cherenkov detector KM3NeT/ORCA \cite{KM3NeT_LoI} 
is currently under construction in the Mediterranean Sea. 
It is optimised for NMO determination by measuring the oscillation probabilities of atmospheric neutrinos in the energy range of $3-30$\,GeV.
The most relevant CP-sensitive energy range is not accessible to ORCA,
necessitating a denser instrumentation to lower the neutrino detection threshold and improve the event reconstruction capabilities.

In this paper, a $\sim 10$ times more-densely instrumented version of ORCA, called Super-ORCA, is considered, aiming at a measurement of the leptonic CP-phase $\delta_{CP}$ from the oscillation pattern of atmospheric neutrinos.
The Super-ORCA detector could also be used as far detector for a future neutrino beam from Protvino \cite{P2O_LoI}. This experimental setup improves significantly the $\delta_{CP}$ sensitivity compared to the measurement with atmospheric neutrinos.
The considered Super-ORCA detector and the estimated event reconstruction capabilities are detailed in Sec.~\ref{sec:SuperORCA_detector}.
The sensitivity calculation methodology and the relevant systematic uncertainties are described in Sec.~\ref{sec:sens_calc}.
The expected sensitivity to measure $\delta_{CP}$ with atmospheric neutrinos is presented in Sec.~\ref{sec:sens_atmo},  
and the $\delta_{CP}$ sensitivity using a neutrino beam from Protvino is discussed in Sec.~\ref{sec:sens_beam}.
A brief conclusion is given in Sec.~\ref{sec:conclusions}.

\vspace{-0.1cm}
\section{Super-ORCA detector and its detector performance}
\label{sec:SuperORCA_detector}
A 10 times denser instrumentation than that realised in the KM3NeT/ORCA detector \cite{KM3NeT_LoI} with a fiducial volume
of 4\,Mton is assumed for Super-ORCA.
This instrumentation density corresponds to 115'000 3-inch PMTs per Mton.\footnote{
For this instrumentation density, `shading' has a few-percent effect on the number of detected Cherenkov photons assuming KM3NeT/ORCA-like technology \cite{KM3NeT_LoI} is used, i.e.\ 31 PMTs in 17-inch glass spheres attached to vertical cables.}
For comparison, this is $\sim1$\%  of the photocathode area density of Super-Kamiokande \cite{SKflavourID}.
This Super-ORCA configuration is the result of a first optimisation of the instrumentation density with the goal to detect about 100 Cherenkov photons per GeV of deposited energy.
This photon statistics allows to distinguish the Cherenkov signatures from electrons and muons of a few hundreds of MeV, resulting in sufficient $\nu_e$/$\nu_\mu$ separation capabilities for $E_\nu \sim 1$\,GeV.
The different angular profiles of the induced Cherenkov radiation are exploited for $e$/$\mu$ separation, similar to the fuzziness of Cherenkov rings as used in Super-Kamiokande \cite{SKflavourID}.

The expected detector performance of Super-ORCA has been estimated based on full maximum-likelihood event reconstructions applied to a simplified detector response simulation. 
A short description of the detector simulation framework and the event reconstruction is given in \cite{P2O_LoI}.
Here, the focus is set to the resulting detector performance.

\begin{figure}[htpb]
\vspace{-0.2cm}
\centering
\begin{minipage}[c]{0.485\textwidth}
\centering
    {\includegraphics[width=\textwidth]{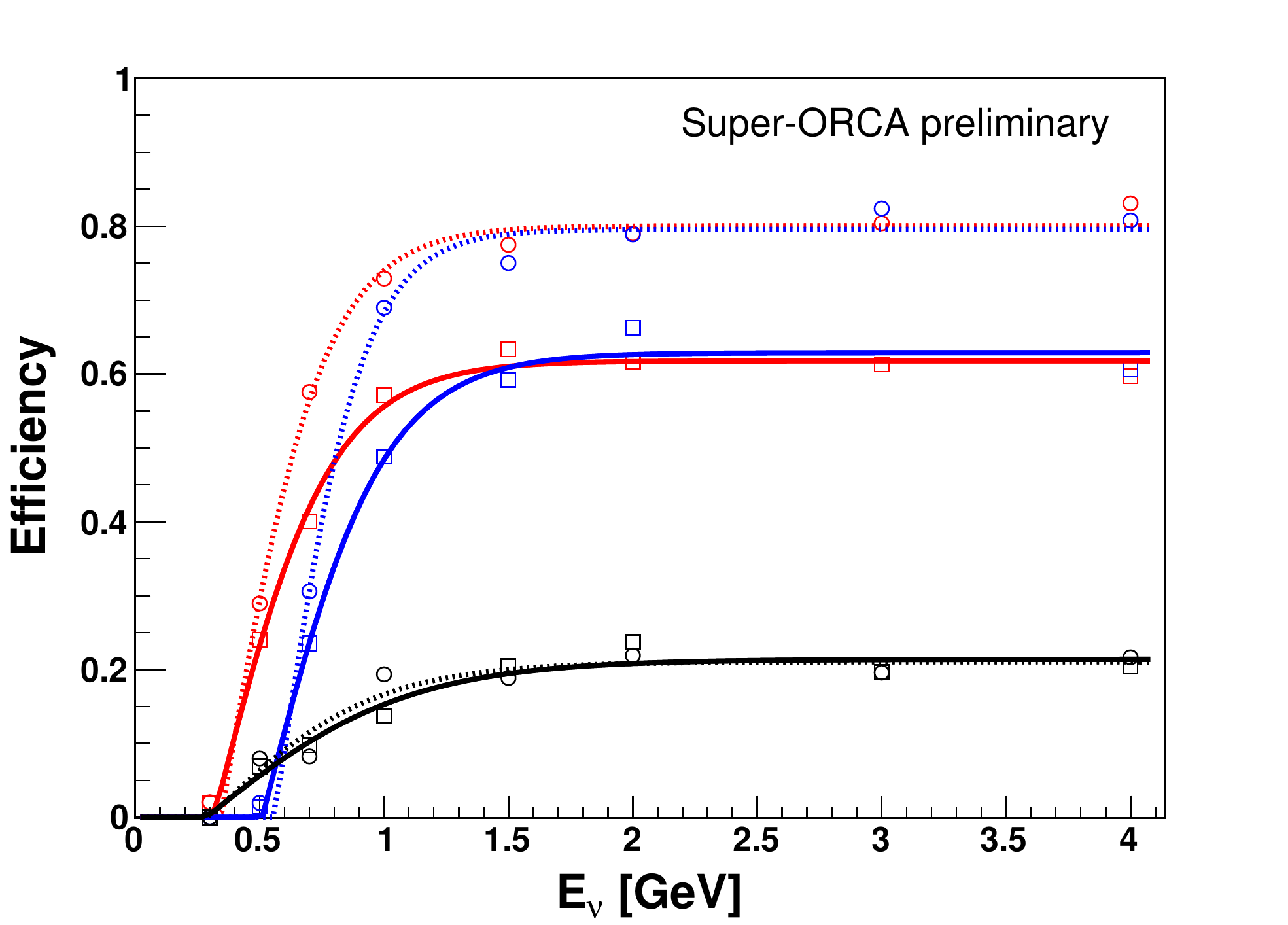}}
\end{minipage}
\hfill
\begin{minipage}[c]{0.485\textwidth}
\centering
    {\includegraphics[width =\textwidth]{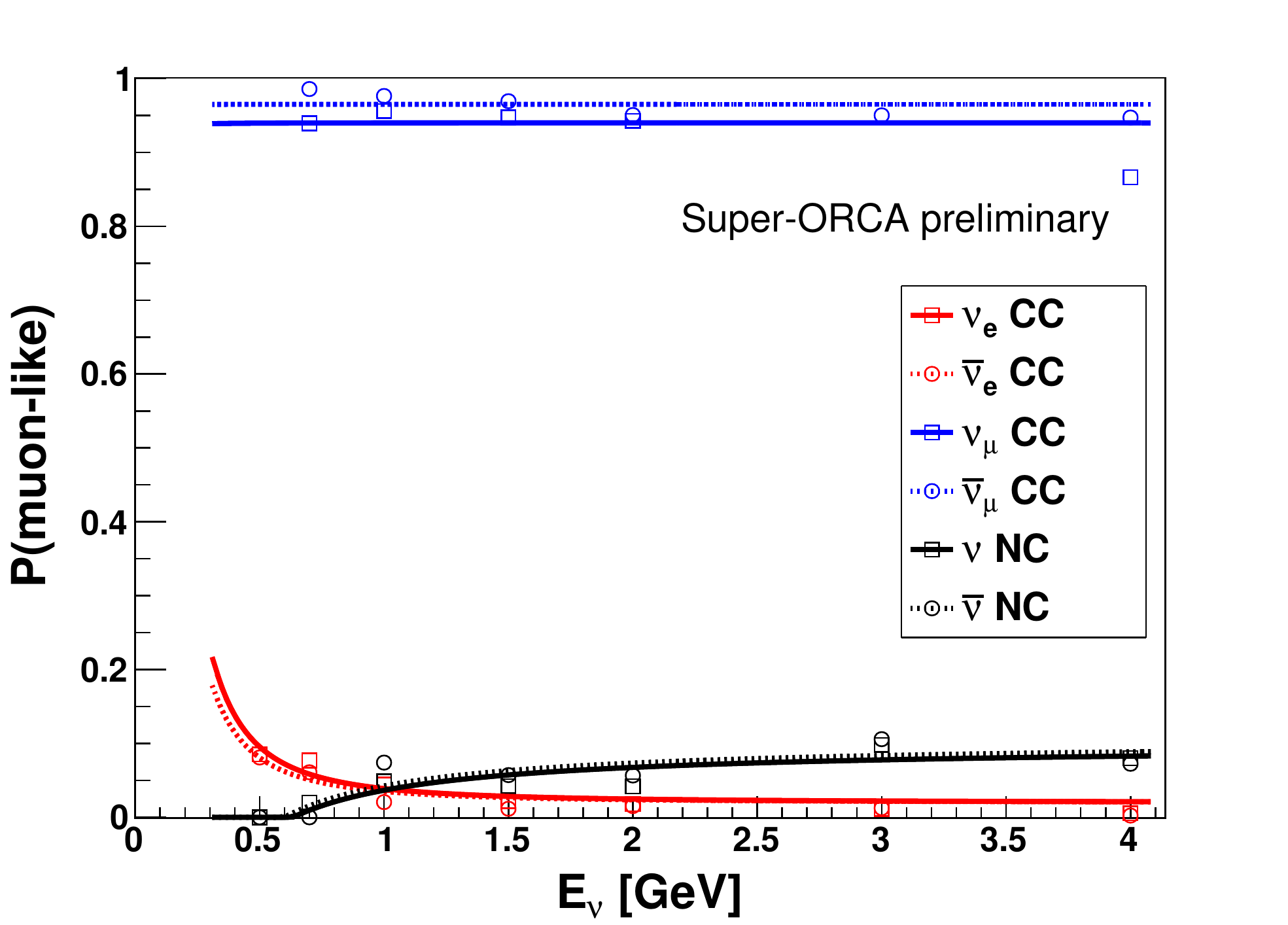}}
\end{minipage}
\vfill
\begin{minipage}[c]{0.485\textwidth}
\centering
    {\includegraphics[width=\textwidth]{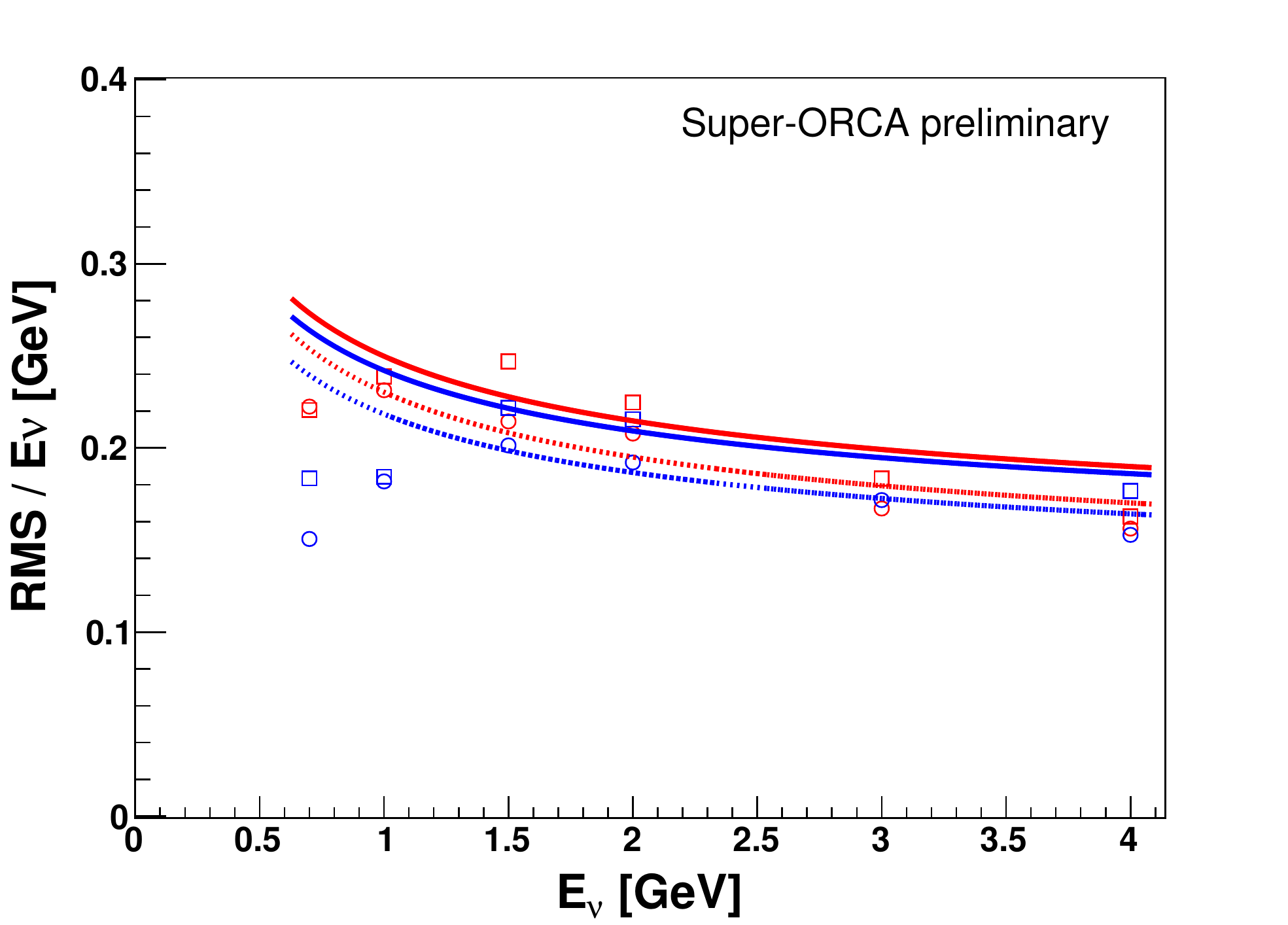}}
\end{minipage}
\hfill
\begin{minipage}[c]{0.485\textwidth}
\centering
    {\includegraphics[width =\textwidth]{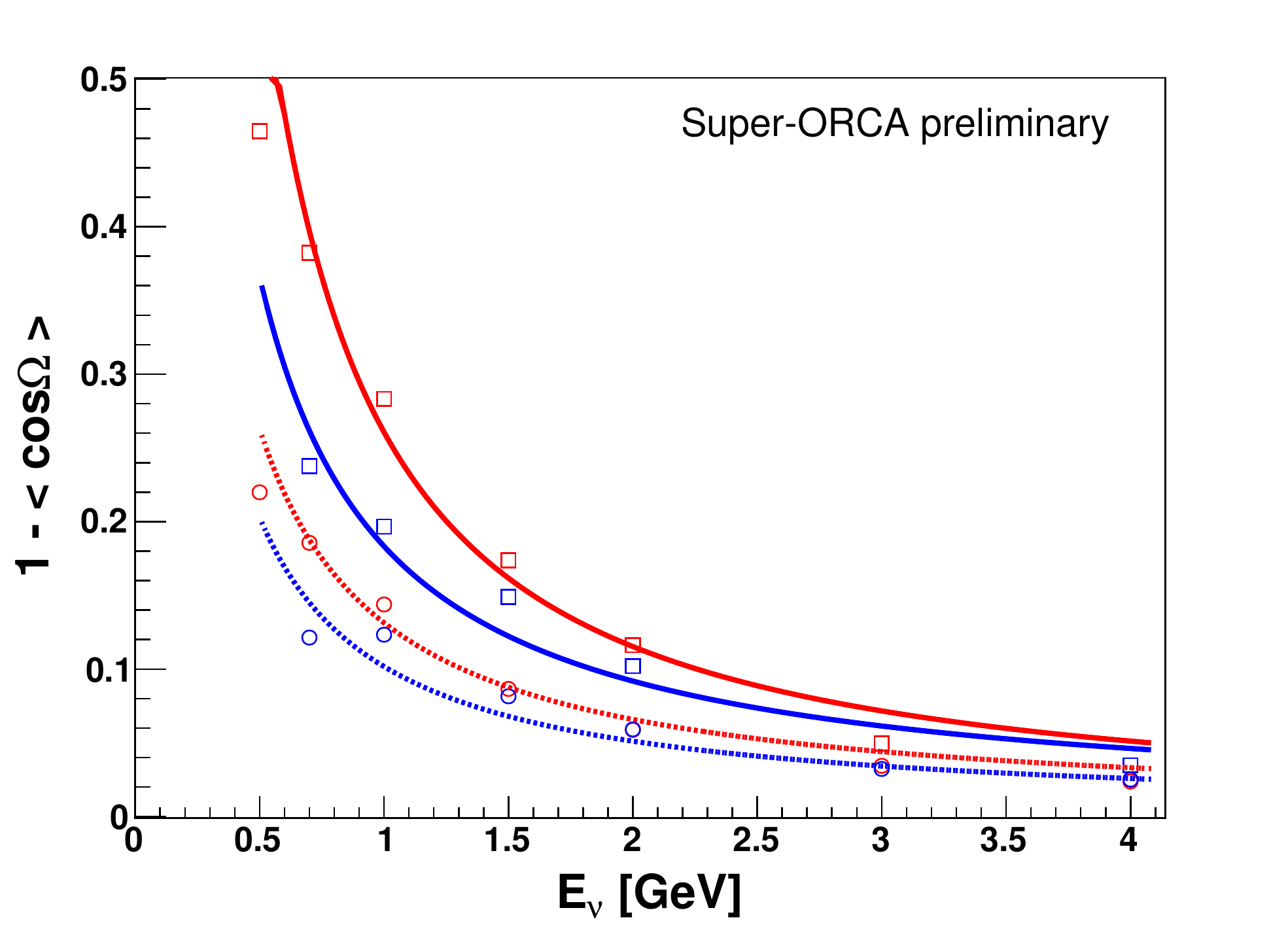}}
\end{minipage}
\caption{
Super-ORCA detector performance derived from simulations for $\nu_e$~CC (red squares), $\bar{\nu}_e$~CC (red circles), $\nu_\mu$~CC (blue squares), $\bar{\nu}_\mu$~CC (blue circles), $\nu$~NC (black squares) and $\bar{\nu}$~CC (black circles) as a function of the true neutrino energy $E_\nu$.
The parametrised detector response functions are shown as solid (dashed) lines for $\nu$ ($\bar{\nu}$).
Top left: neutrino detection efficiency, including event selection.
Top right: probability $P$ to classify an event as muon-like. The probability for classification as electron-like is then $1-P$, as two classes are considered.
Bottom left: relative neutrino energy resolution in terms of RMS of the reconstructed energy distribution for a given true $E_\nu$ divided by $E_\nu$.
Bottom right: neutrino direction resolution quantified as $1-\langle \cos\Omega \rangle$, where $\langle \cos\Omega \rangle$ is average cosine of the angle $\Omega$ between the neutrino direction and the reconstructed direction. A value of $1-\langle \cos\Omega \rangle = 0.2$ corresponds to $\sim 37^\circ$.
}
\label{fig:detector_performance}
\end{figure}

\vspace{-0.1cm}
\paragraph{Detector performance\\}
\label{sec:detector_performance}
The Super-ORCA detector performance is summarised in Fig.~\ref{fig:detector_performance}.
In general, the detector performance is better for 
events induced by $\bar{\nu}$ charged-current (CC) interactions than for $\nu$~CC interactions
due to the smaller average interaction inelasticity (Bjorken $y$) for $\bar{\nu}$~CC.
The outgoing $e$/$\mu$ produce more Cherenkov light than the hadrons for the same kinetic energy, 
and in addtion the Cherenkov signatures from $e$/$\mu$ show much less event-by-event fluctuations than that of hadrons \cite{intrinsicLimits}. 

The energy threshold for neutrino detection (Fig~\ref{fig:detector_performance} top left) is about $\sim 0.5$\,GeV for $\nuan_e$~CC events and about $0.2$\,GeV larger for $\nuan_\mu$~CC due to the higher energy threshold for Cherenkov radiation for muons compared to electrons.
The detection efficiency saturates at a value of $\sim 85$\% for $\bar{\nu}_{e,\mu}$~CC and $\sim 65$\% for $\nu_{e,\mu}$~CC. 
With the assumed fiducial volume of 4\,Mton,
the resulting effective volume of the detector is $\sim 3.4$/$2.6$\,Mton for $\bar{\nu}_{e,\mu}$/$\nu_{e,\mu}$~CC events.
NC events show a lower efficiency of $\sim 20$\%, as they are partly suppressed due to the applied event selection criteria favouring clear e/$\mu$-like Cherenkov cone and suppressing events with several Cherenkov cones from different hadrons.

The probability to classify an event as {\it muon-like} (Fig~\ref{fig:detector_performance} top right) is about $95$\% for $\nuan_\mu$~CC events and below $5$\% for $\nuan_e$~CC for neutrino energies above $1$\,GeV,
so that $\sim 95$\% of the $\nuan_e$~CC events are classified as {\it electron-like}.
Most NC events are classified as electron-like,
while $\sim 10$\% of them are classified as muon-like.

The neutrino energy resolution (Fig~\ref{fig:detector_performance} bottom left) is better than $\sim 20-25$\% for $E_\nu > 1$\,GeV and is dominated by fluctuations in the number of emitted photons in the hadronic shower \cite{intrinsicLimits}.
The energy resolution improves for very small neutrino energies ($E_\nu < 1$\,GeV) which is a result of the event selection that introduces a bias towards low Bjorken-y events in the energy regime of the detection efficiency turn-on (see Fig~\ref{fig:detector_performance} top left), as discussed in \cite{intrinsicLimits}.

The neutrino direction resolution (Fig~\ref{fig:detector_performance} bottom right) is $\sim 44^{\circ}$/$31^{\circ}$/$36^{\circ}$/$28\,^{\circ}$ for $\nu_e$/$\bar{\nu}_e$/$\nu_\mu$/$\bar{\nu}_\mu$~CC events with $E_\nu = 1$\,GeV, and improves for higher energies.
The resolution on the outgoing lepton is a few degree, 
and the resolution on the neutrino direction is dominated by the intrinsic neutrino-lepton scattering angle \cite{intrinsicLimits}.

\vspace{-0.2cm}
\section{Sensitivity calculation methodology}
\label{sec:sens_calc}
\vspace{-0.05cm}
The sensitivity to measure $\delta_{CP}$ is calculated following an {\it Asimov dataset} approach with a $\chi^2$-minimisation and simultaneously fitting several nuisance parameters.
A similar procedure is applied in \cite{KM3NeT_LoI, P2O_LoI, Bruchner2018} and is further described therein.

A set of parametrised detector response functions is derived from the results presented in Sec.~\ref{sec:detector_performance}. These functions are shown as lines in Fig.~\ref{fig:detector_performance}.\footnote{As $\nuan_{\tau}$~CC events are less relevant for measuring $\delta_{CP}$  with atmospheric neutrinos, no dedicated $\nuan_{\tau}$~CC simulations have been performed. The detector response for $\nuan_{\tau}$~CC events is assumed to be identical to that of NC events taking the $\nuan_{\tau}$~CC interaction cross sections into account.}
The atmospheric neutrino fluxes are modelled using the HKKM2014 simulations \cite{Honda},
and the neutrino-nucleon cross sections are modelled using GENIE \cite{GENIE} predictions for an oxygen nucleus and two protons.
Oscillation probabilities are computed with OscProb \cite{OscProb}.

\begin{figure}[htpb]
\vspace{-0.2cm}
\centering
\begin{minipage}[l]{0.65\textwidth}
\begin{minipage}[c]{0.48\textwidth}
\centering
    {\includegraphics[width=\textwidth]{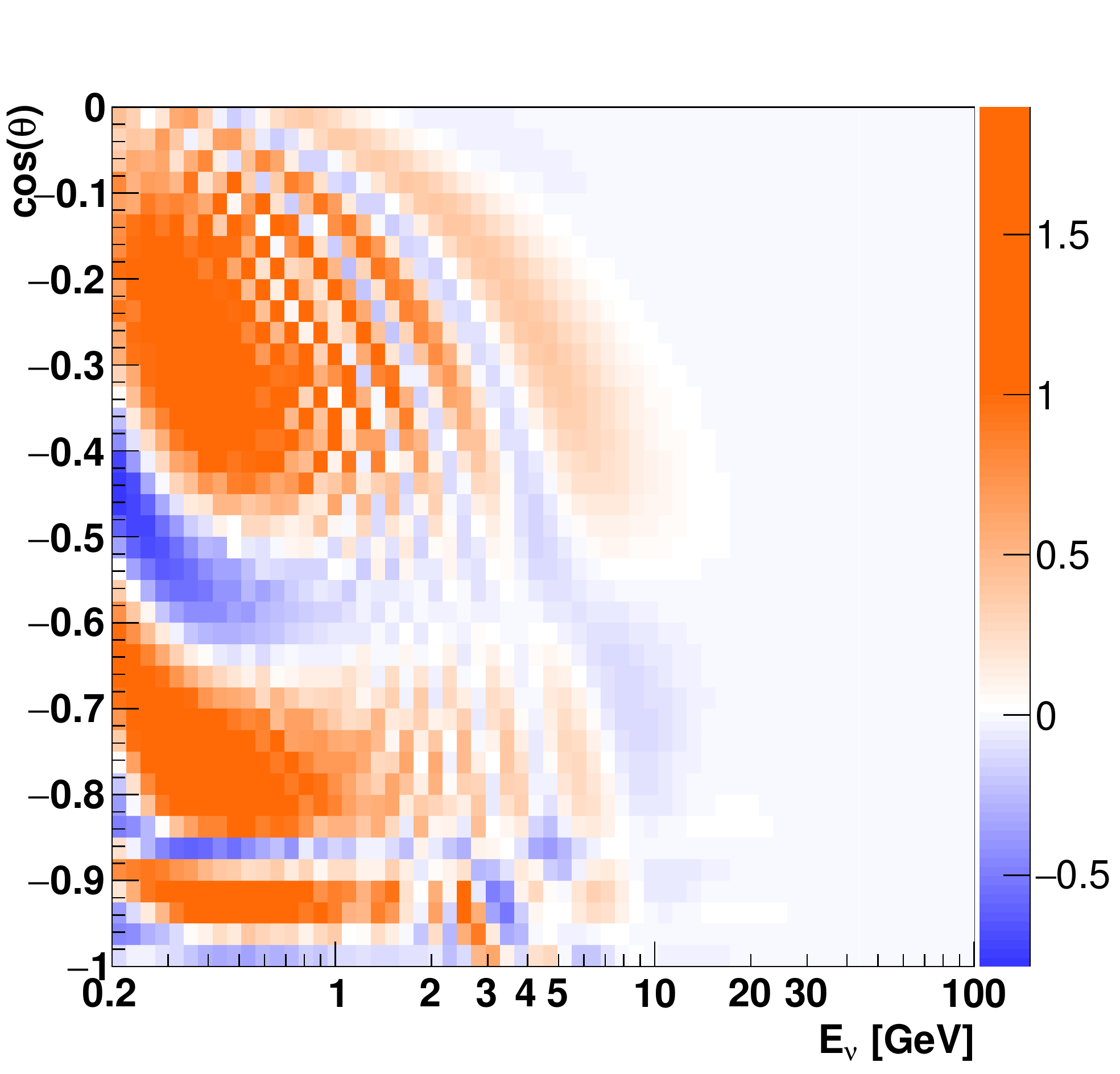}}
\end{minipage}
\hfill
\begin{minipage}[c]{0.48\textwidth}
\centering
    {\includegraphics[width =\textwidth]{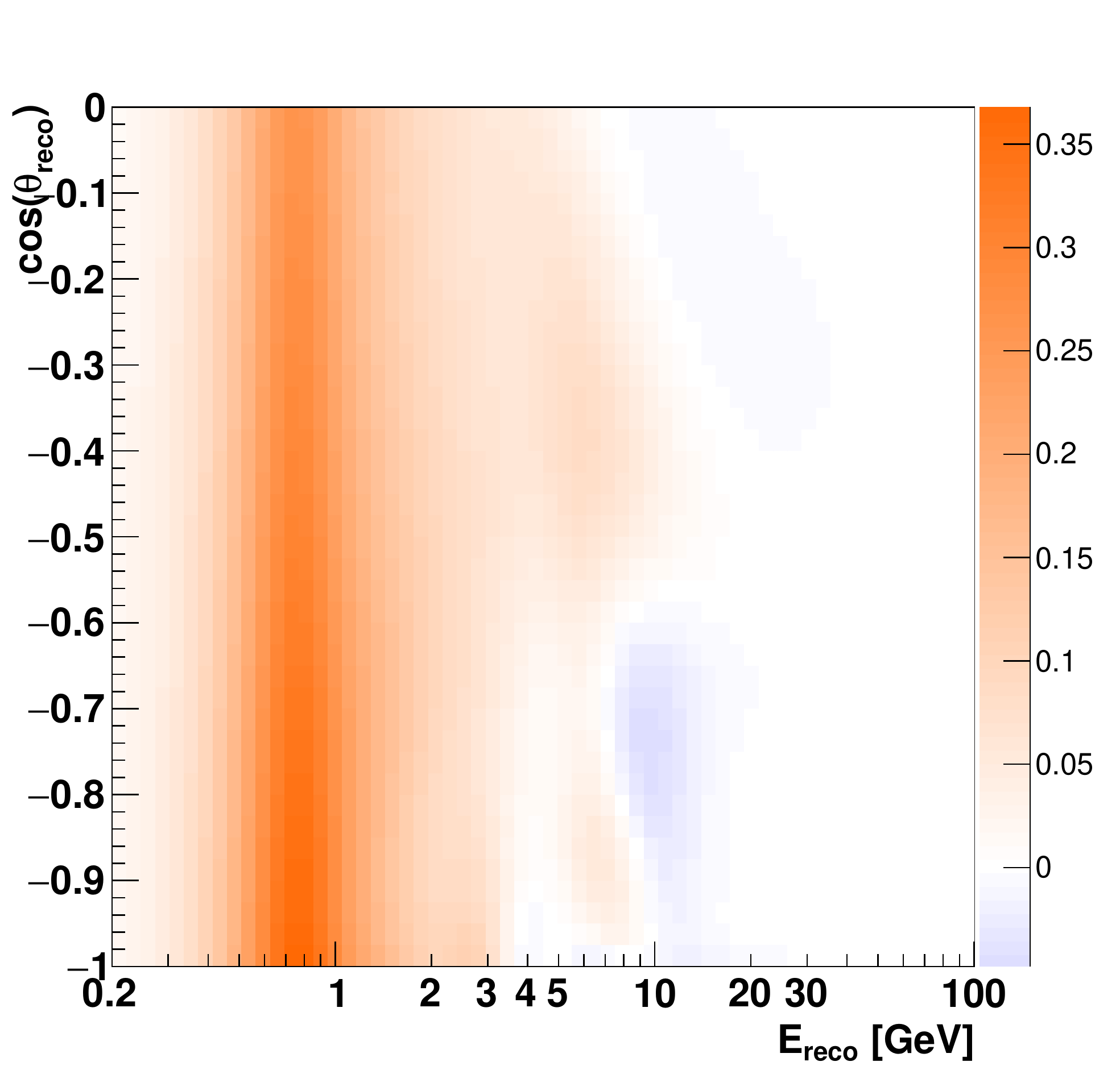}}
\end{minipage}
\vfill
\begin{minipage}[c]{0.48\textwidth}
\centering
    {\includegraphics[width=\textwidth]{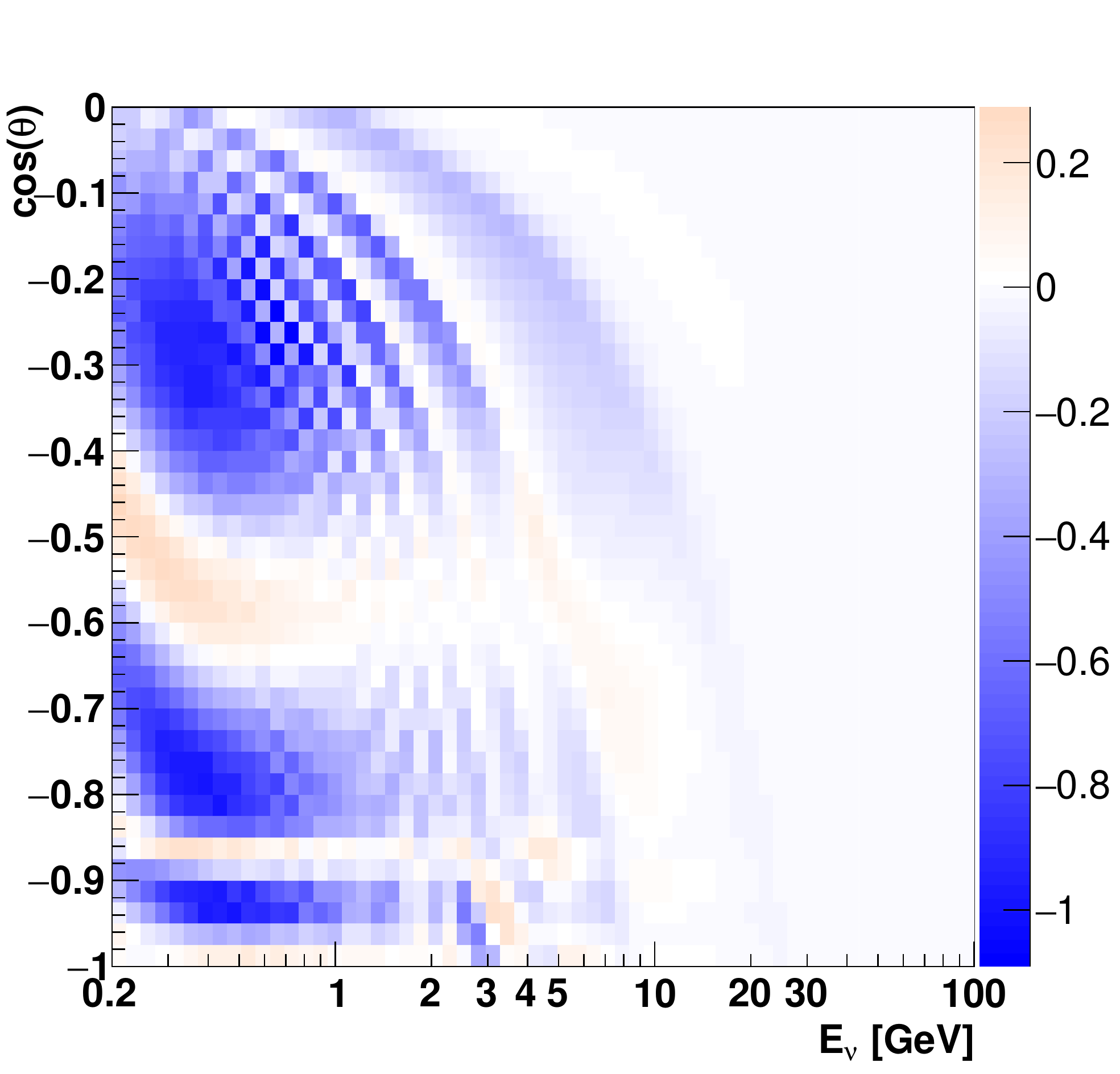}}
\end{minipage}
\hfill
\begin{minipage}[c]{0.48\textwidth}
\centering
    {\includegraphics[width =\textwidth]{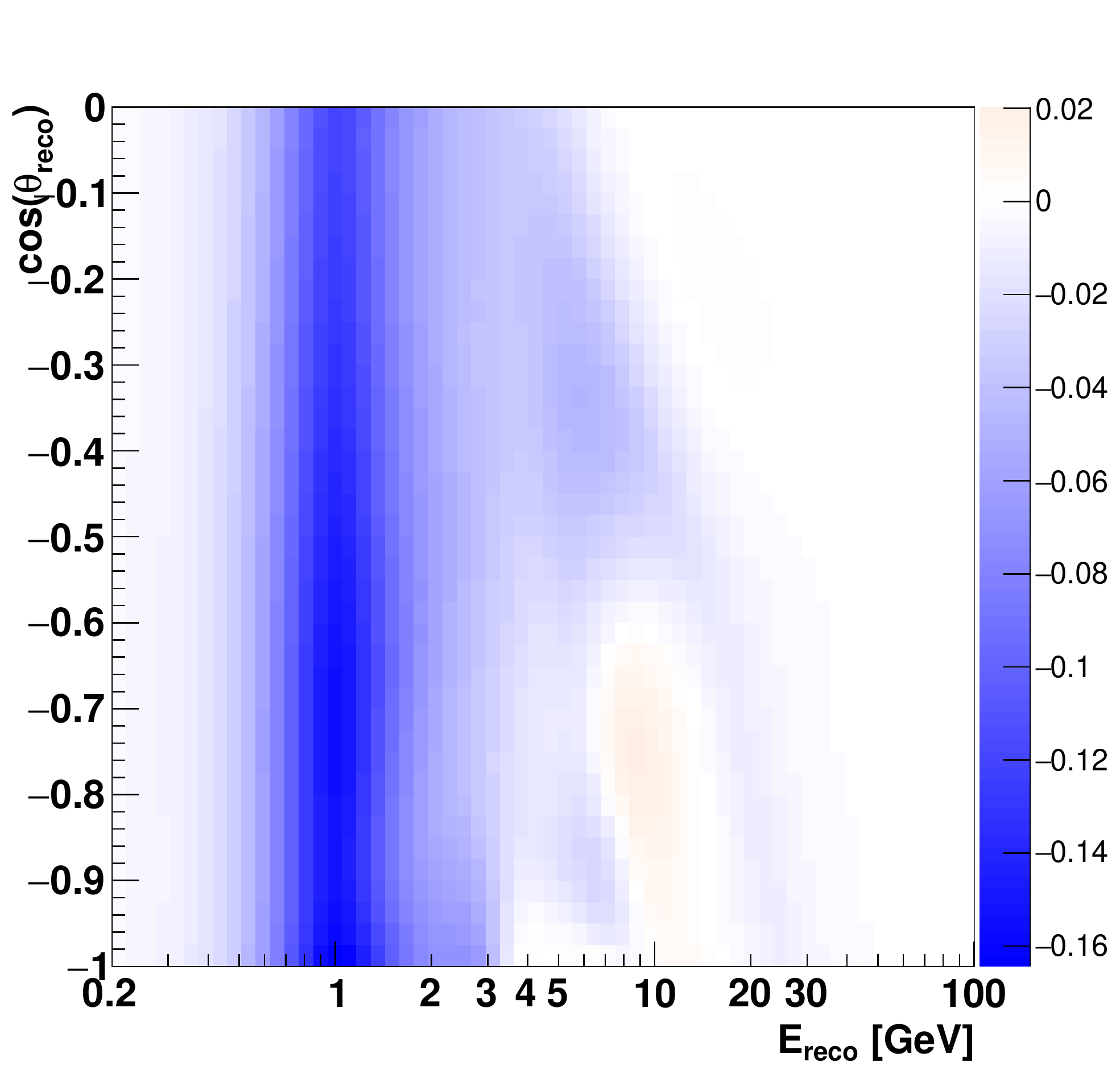}}
\end{minipage}
\end{minipage}
\begin{minipage}[c]{0.03\textwidth}
~
\end{minipage}
\begin{minipage}[c]{0.3\textwidth}
\caption{
Asymmetry $\chi$ (as defined by Eq.~\ref{eq:chi}) between the number of $\nu$+$\bar{\nu}$~CC events expected for $\delta_{CP}=0$ and $\delta_{CP}=3/2\pi$ as a function of $E_\nu$ and $\cos(\theta_z)$. The top/bottom plot applies to electron/muon neutrinos. The left/right plot is before/after applying the detector response (as given in Fig.~\ref{fig:detector_performance}).
In total, $\sim 160'000$~$\nu_e$+$\bar{\nu}_e$~CC and $\sim 140'000$~$\nu_\mu$+$\bar{\nu}_\mu$~CC events per year are observed.
}
\label{fig:asymmetries}
\end{minipage}
\end{figure}

The general idea is to calculate the median significance to reject a test $\delta_{CP}^{test}$ assuming a true value $\delta_{CP}^{true}$. 
The statistical significance to distinguish between two $\delta_{CP}$ values is calculated from the number $N$ of events in bins of reconstructed energy and cosine of the reconstructed zenith angle $\theta_z$.
The statistical significance $\chi^2$ is computed from the event number asymmetry $\chi$ in each bin:
\begin{equation}
\chi = \left ( N_{\delta_{CP}^{test}} - N_{\delta_{CP}^{true}} \right ) /  \sqrt{N_{\delta_{CP}^{true}}}.
\label{eq:chi}
\end{equation}
The total significance is then given by the sum of $\chi^2$ over all bins of the electron-like as well as the muon-like event histograms.
For illustration purposes, the asymmetry $\chi$ to distinguish between $\delta_{CP}=0$ and $\delta_{CP}=3/2\pi$ is shown in Fig.~\ref{fig:asymmetries} for $\nuan_e$~CC and $\nuan_\mu$~CC, separately.
Its pronounced pattern illustrates the achievable separation power.
After detector smearing some $\delta_{CP}$ sensitivity remains, 
however the fine-grained pattern is blurred by the detector response.
Most of the sensitivity is contributed by $\nuan_e$~CC events.
The sign of the asymmetry $\chi$ (Eq.~\ref{eq:chi}) is opposite for $\nuan_e$~CC and $\nuan_\mu$~CC, underlining the importance of good $\nuan_e$ vs $\nuan_\mu$ separation.
The most relevant energy range is $\sim 1$\,GeV for $\nuan_e$~CC as well as for $\nuan_\mu$~CC, however, with a small shift in reconstructed energy, indicating the importance of a possible skew in the energy measurement scale.
The sign of the asymmetry $\chi$ is also opposite for $\nu$ than for $\bar{\nu}$,
which is taken statistically into account due to different interaction rates and different detector responses for $\nu$ and $\bar{\nu}$ (in particular detection efficiency, Fig.~\ref{fig:detector_performance} top left).
Increasing the $\delta_{CP}$ sensitivity due to exploiting the measured inelasticities is not yet included, but is under study. 

Several systematic uncertainties considering the atmospheric neutrino flux, oscillation parameters, interaction cross sections and detector-related energy measurement scales are considered in the fit.
The complete list of parameters together with their assumed true values and Gaussian priors is given in Table~\ref{tab:syst}.

\begin{table}
\begin{minipage}[c]{0.39\textwidth}
\begin{tabular}{ | c | c | c | c | c | c | c | c | }
  \hline  \hline
Parameter  & value/prior \\ \hline
$\theta_{13}$  &   $8.51^\circ\pm0.15^\circ$ \\ 
$\theta_{23}$  &   $45.0^\circ$ (free) \\
$\Delta M^2$ $[10^{-3}$ eV$^2]$ &  $2.5 \pm 0.05$ \\ 
  \hline
overall norm &     $10$\% \\ 
NC norm   &     $10$\% \\ 
$\nu_\tau$ norm  &     $10$\% \\ 
$\nu_e/\nu_\mu$ skew  &     $10$\% \\ 
$\nu/{\bar \nu}$ skew  &     $3$\% \\ 
flux E-tilt &    $0.05$ \\ 
flux $\cos(\theta_z)$-tilt &    $2$\% \\ 
  \hline
ParticleID skew    &     $10$\% \\ 
$E_{\textrm{scale}}$ overall        & $3$\% \\ 
$E_{\textrm{scale}}$ e/$\mu$ skew   & $3$\% \\ 
$E_{\textrm{scale}}$ had/e skew 		& $3$\% \\ 
$E_{\textrm{scale}}$ $\nu/{\bar \nu}$ skew & $3$\% \\ 
$E_{\textrm{scale}}$ $\cos(\theta_z)$-tilt & $3$\% \\ 
  \hline \hline
\end{tabular}
\end{minipage}
\begin{minipage}[c]{0.02\textwidth}
\end{minipage}
\begin{minipage}[c]{0.59\textwidth}
\caption{
List of relevant parameters and their uncertainties for neutrino oscillations, and parameters for neutrino fluxes, neutrino interactions and detector-related systematics, as well as their priors.\newline
The considered systematic uncertainties include an overall normalisation, 
an independent normalisation for NC as well as $\nu_\tau$~CC,
a skew between $\nu_e$ and $\nu_\mu$,
a skew between $\nu$ and $\bar \nu$,
an energy-dependent tilt (spectral index) as well as a $\cos(\theta_z)$-dependent tilt in the atmospheric neutrino fluxes, 
a skew in the event identification as e-like or $\mu$-like, 
and five different energy scale parameters related to systematic uncertainties in the energy measurement.
The energy scale parameters are an overall energy scale, 
and two skew parameters that allow separate energy scales for $\nu_e$, $\nu_\mu$ and hadronic channels (NC and $\nu_\tau$),
an energy scale skew between $\nu$ and $\bar \nu$, 
and an energy scale skew between up/horizontal events (assumed as $\cos(\theta_z)$-dependent tilt). \newline
The choice of priors for the 
uncertainties related to atmospheric neutrino fluxes are motivated by \cite{Barr_uncertainties}.
The choice of values for the other priors is motivated by sensitivity studies performed by KM3NeT/ORCA \cite{KM3NeT_LoI} and by long-baseline experiments \cite{DUNE}.
}
\label{tab:syst}
\end{minipage}
\end{table}

\vspace{-0.25cm}
\section{Sensitivity to $\delta_{CP}$ using atmospheric neutrinos }
\label{sec:sens_atmo}
\begin{figure}[htpb]
\vspace{-0.15cm}
\centering
\begin{minipage}[c]{0.51\textwidth}
\centering
    {\includegraphics[width =\textwidth]{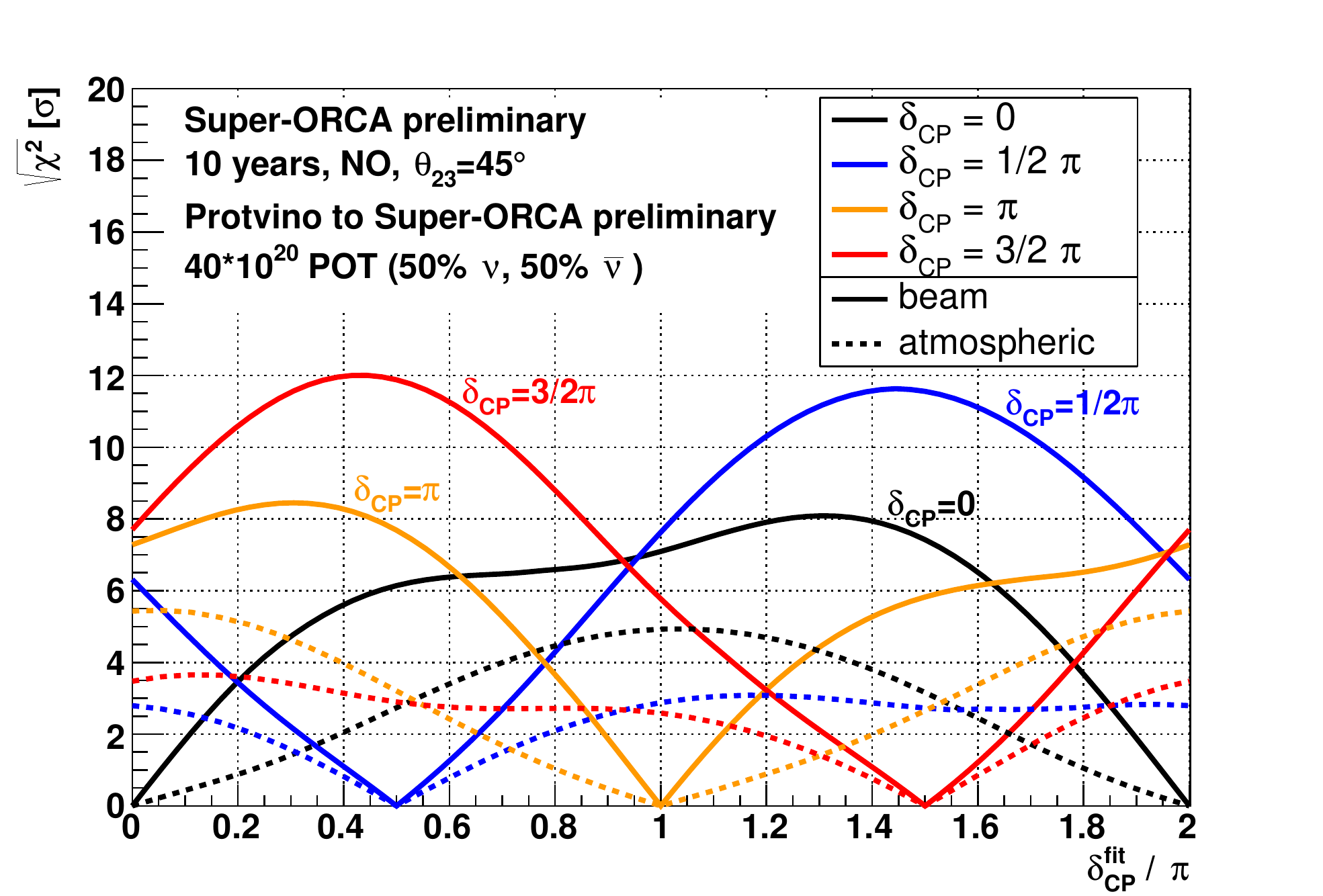}}
\end{minipage}
\hfill
\begin{minipage}[c]{0.46\textwidth}
 \caption{Sensitivity to exclude certain values of $\delta_{CP}$ with Super-ORCA after 10 years of data taking as a function of the tested $\delta_{CP}^{\rm fit}$ for 4 example values of true $\delta_{CP}$ using atmospheric neutrinos (dashed lines).
Normal mass ordering is assumed.
The corresponding sensitivity for Super-ORCA using a 450\,kW beam from Protvino with 5 years in neutrino mode and 5 years in antineutrino mode is shown for comparison (solid lines). 
}
\label{fig:atmoANDbeam_fourChi2curves}
\end{minipage}
\end{figure}

\vspace{-0.05cm}
The expected sensitivity to distinguish between different $\delta_{CP}$ values with Super-ORCA after 10~years of data from atmospheric neutrinos (dashed lines) is shown in Fig.~\ref{fig:atmoANDbeam_fourChi2curves}.
The largest sensitivity is achieved between $\delta_{CP}=0$ and $\delta_{CP}=\pi$ with $5\,\sigma$.
About 63\%  (72\%) of the $\delta_{CP}$ values can be disfavoured with $\geq 2 \sigma$ for true $\delta_{CP}=0$ and $\delta_{CP}=\pi$ ($\delta_{CP}=\pi/2$ and $\delta_{CP}=3/2\pi$).
$\delta_{CP}=0$ and $\delta_{CP}=\pi$ ($\delta_{CP}=\pi/2$ and $\delta_{CP}=3/2\pi$).
The $1\sigma$ resolution is about $38^\circ$ ($23^\circ$) for true $\delta_{CP}=0$ and $\delta_{CP}=\pi$ ($\delta_{CP}=\pi/2$ and $\delta_{CP}=3/2\pi$).

\begin{figure}[htpb]
\vspace{-0.1cm}
\centering
\begin{minipage}[c]{0.495\textwidth}
\centering
	\includegraphics[width=\textwidth]{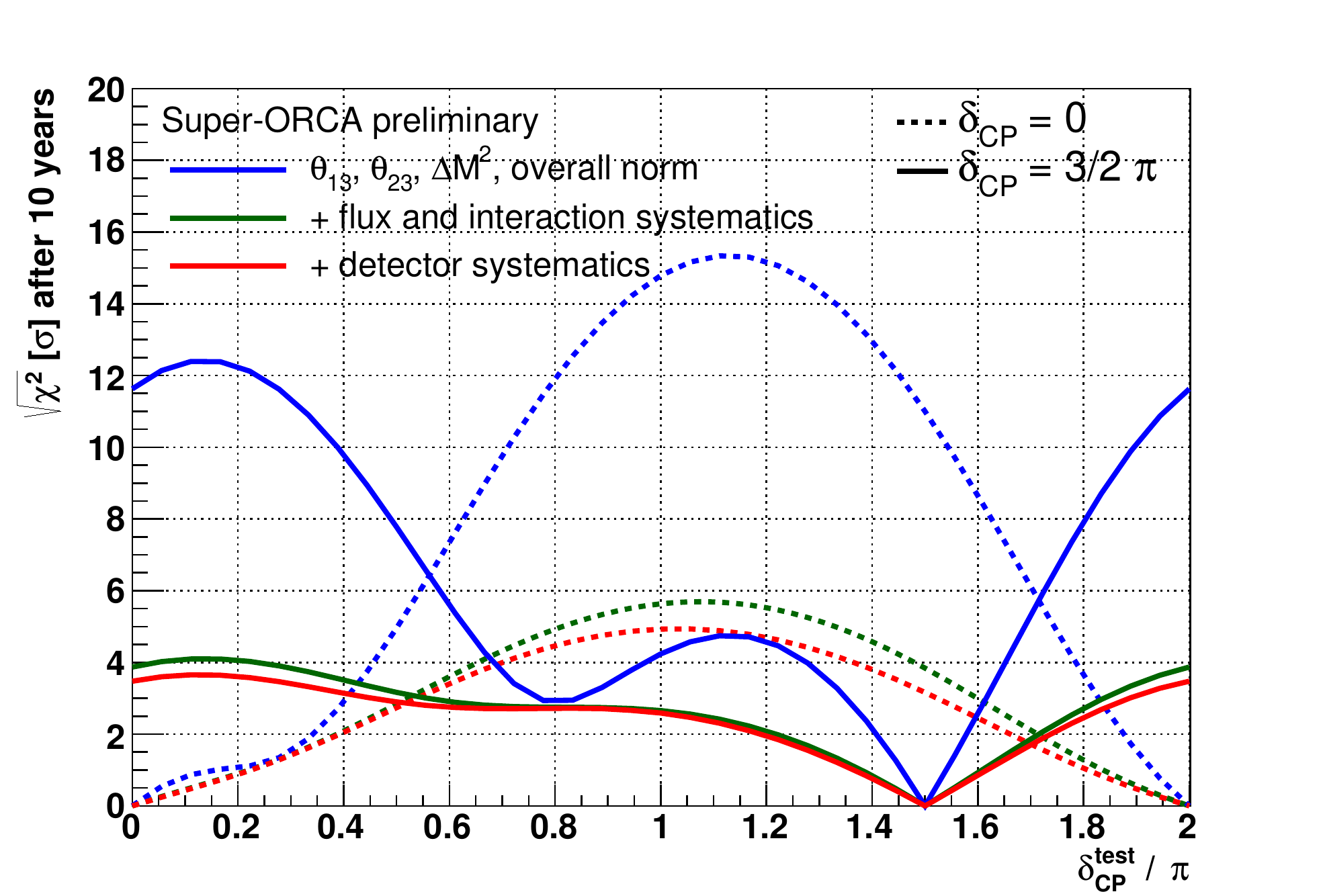}
\end{minipage}
\hfill
\begin{minipage}[c]{0.495\textwidth}
\centering
	\includegraphics[width=\linewidth]{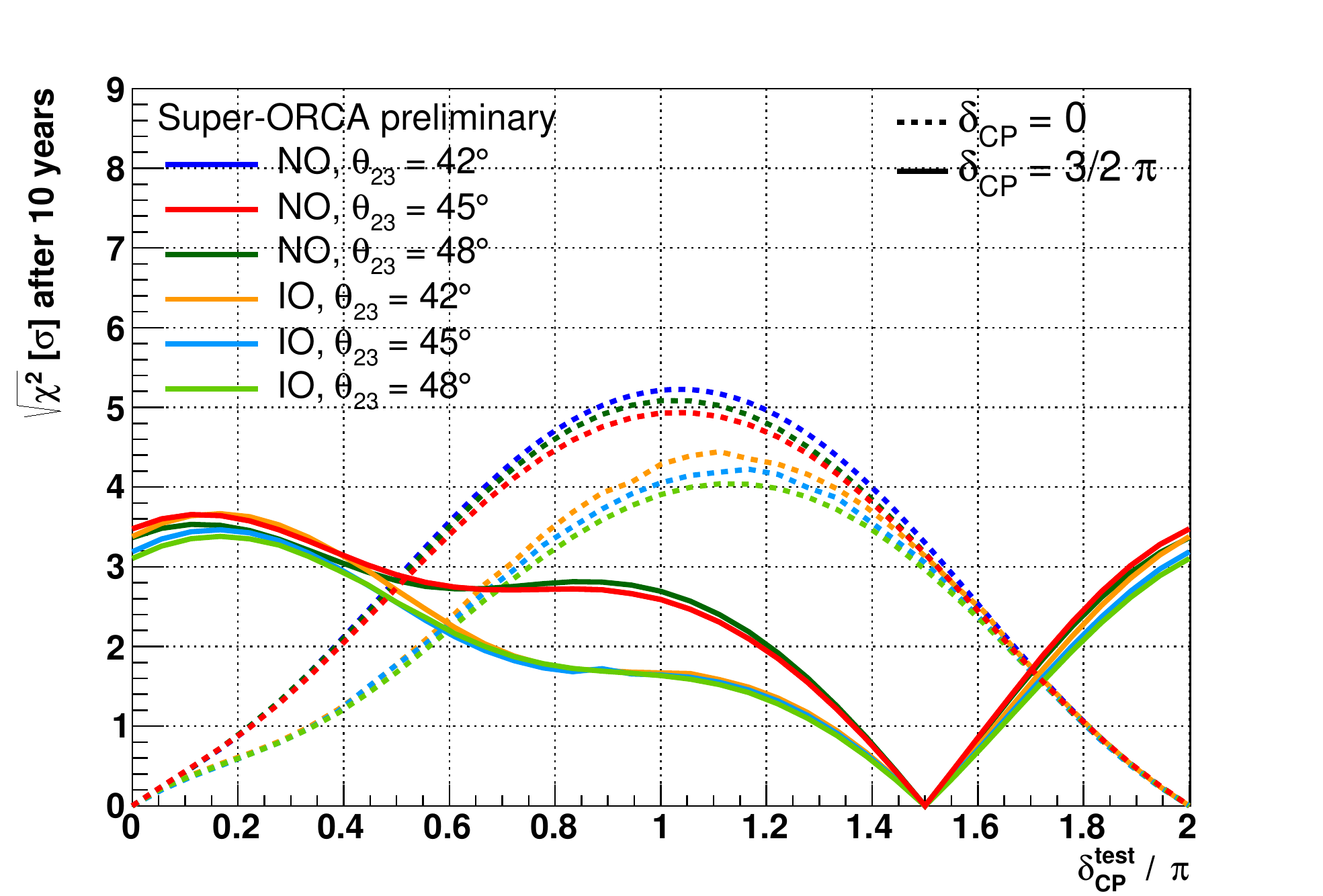}
\end{minipage}
\caption{
Left: 
sensitivity to exclude certain values of $\delta_{CP}$ with Super-ORCA after 10 years for $\delta_{CP}=0$ and $\delta_{CP}=3/2\pi$,
including three different sets of systematic uncertainties: 1) only $\theta_{13}$, $\theta_{23}$, $\Delta M^2$ and an overall normalisation, 2) adding flux- and interaction-related systematics and 3) adding detector-related systematics.
Right:
$\delta_{CP}$ sensitivity for different values of $\theta_{23}$ and for normal/inverted (NO/IO) mass ordering. 
\label{fig:atmo_sensitivity_systematics_theta23NMH}
}
\end{figure}

\vspace{-0.1cm}
\paragraph{Effect of systematics\\}
The effect of different classes of systematics can be inferred from Fig.~\ref{fig:atmo_sensitivity_systematics_theta23NMH} (left). 
Neutrino flux and interaction systematics have the largest effect on the $\delta_{CP}$ sensitivity, while including the detector-related systematics reduces the sensitivity only mildly.

The effect of the unknown value of $\theta_{23}$ and the NMO is shown in Fig.~\ref{fig:atmo_sensitivity_systematics_theta23NMH} (right).
A larger  $\delta_{CP}$ sensitivity is achieved for NO compared to IO.
The  $\delta_{CP}$ sensitivity depends weakly on the true value of $\theta_{23}$.

\vspace{-0.1cm}

\vspace{-0.15cm}
\section{Sensitivity to $\delta_{CP}$ using a neutrino beam from Protvino}
\label{sec:sens_beam}
\vspace{-0.05cm}
The potential of Super-ORCA and its $\delta_{CP}$ sensitivity can be significantly improved when using a neutrino beam instead of atmospheric neutrinos due to the ability to control the beam polarity ($\nu$ and $\bar \nu$ modes).
A suitable candidate accelerator facility is located in Protvino in the Moscow region, Russia.
The baseline is 2595\,km to the KM3NeT-Fr site where ORCA is located.
This proposed experimental setup is known as P2O \cite{P2O_LoI}.
The design of the main synchrotron of the Protvino accelerator facility potentially allows for operation at a beam power up to 450\,kW.
Further details on the accelerator and the P2O proposal are given in \cite{P2O_LoI}.

\begin{figure}[htpb]
\vspace{-0.1cm}
\centering
\begin{minipage}[c]{0.495\textwidth}
\centering
    {\includegraphics[width=\textwidth]{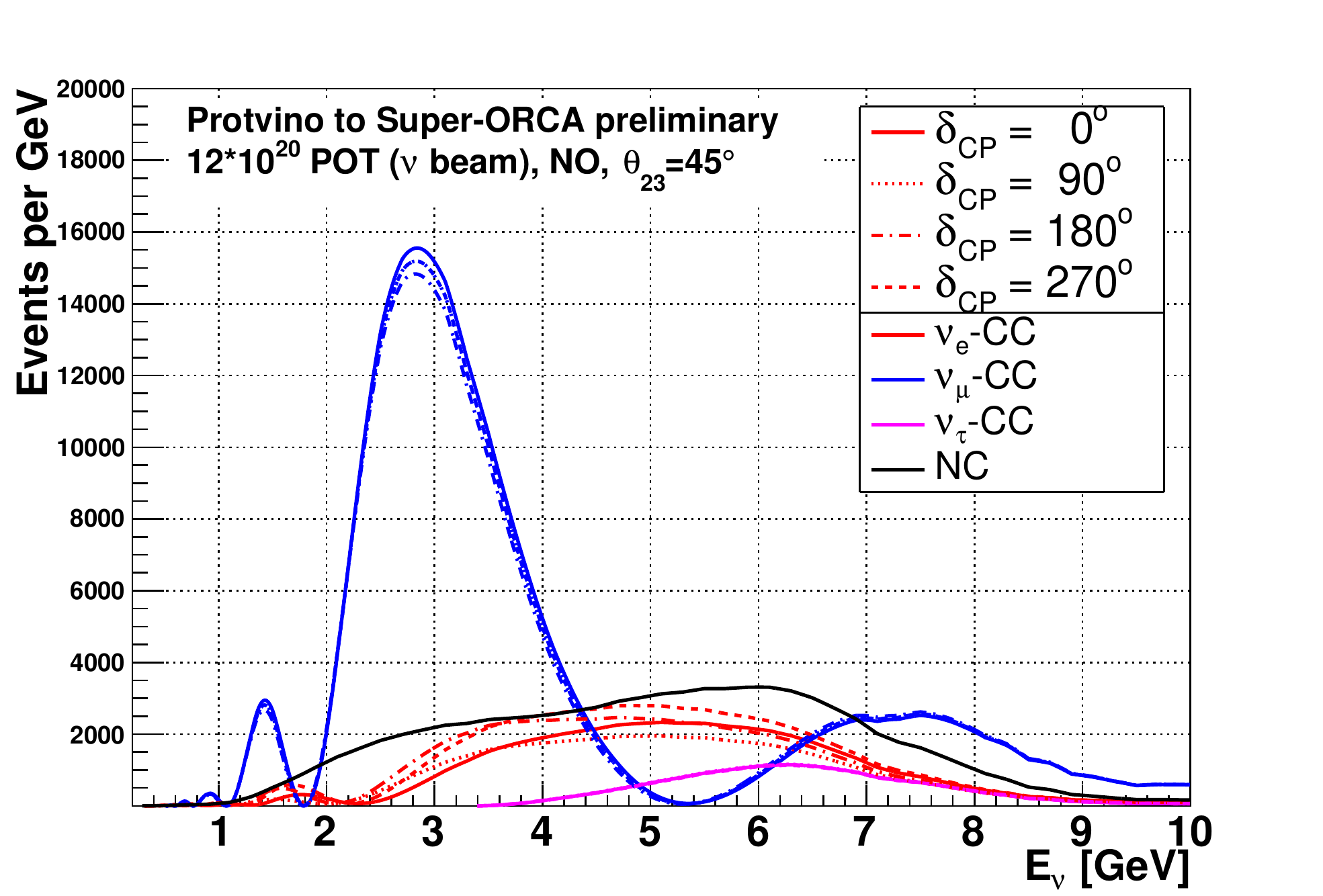}}
\end{minipage}
\hfill
\begin{minipage}[c]{0.495\textwidth}
\centering
    {\includegraphics[width =\textwidth]{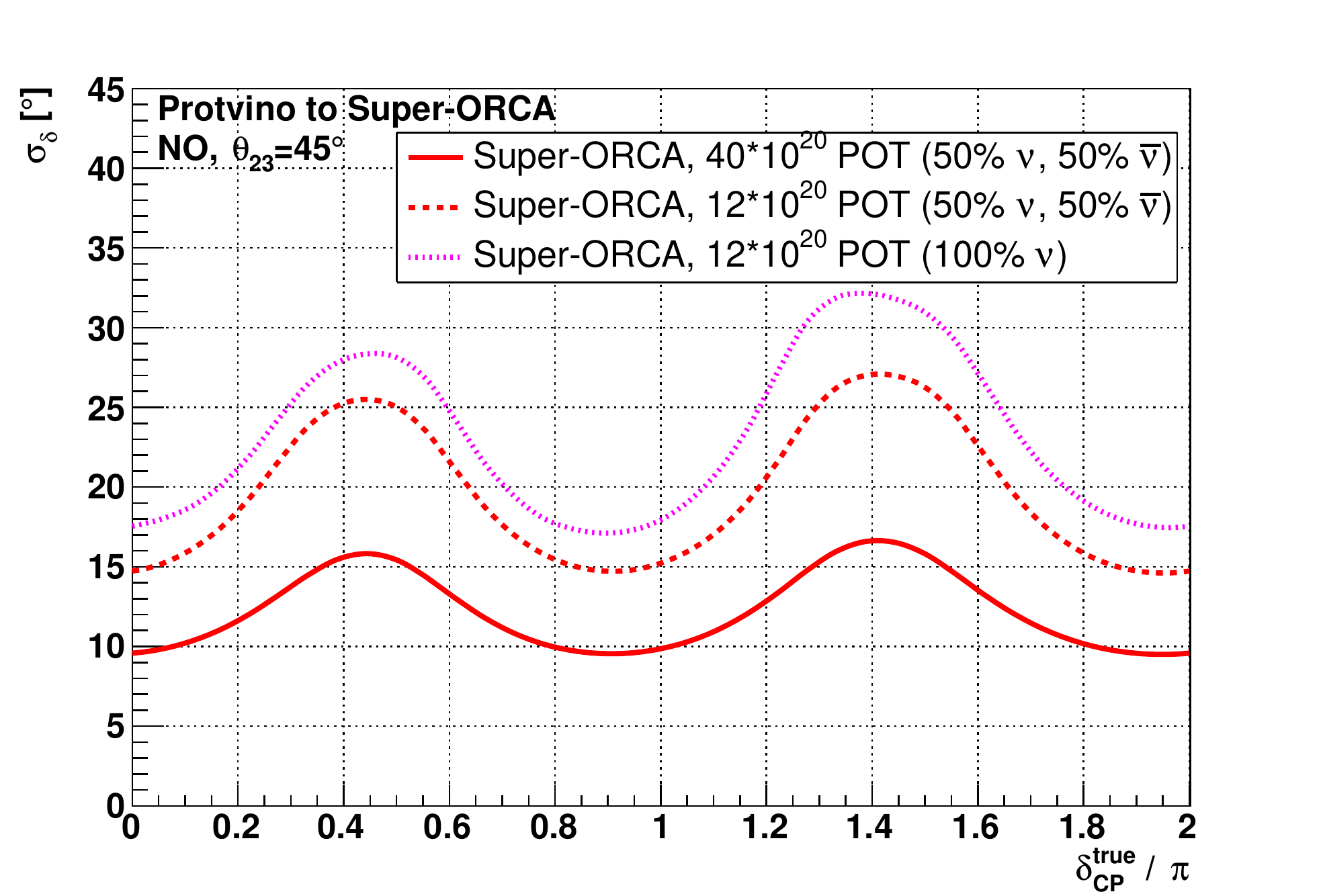}}
\end{minipage}
\caption{
Left:
energy distribution of number of neutrino events detected with Super-ORCA after 3 years of running with a 450\,kW beam from Protvino (neutrino mode) for 4 different values of $\delta_{CP}$ ($0$, $\pi/2$, $\pi$, $3/2\pi$). 
Right \cite{P2O_LoI}:
1$\sigma$-resolution on $\delta_{CP}$ as function of the true $\delta_{CP}$ value for Super-ORCA and the 450\,kW beam operating 3 years  with 100\% $\nu$ beam (dotted line) and 50\%$\nu$/50\%$\bar \nu$ beam (dashed line) and 10 years with 50\%$\nu$/50\%$\bar \nu$ beam (solid line).
Normal neutrino mass ordering and $\theta_{23}=45^\circ$ is assumed for both plots. 
}
\label{fig:beam_Nevents_deltaCPresolution}
\end{figure}

Fig.~\ref{fig:beam_Nevents_deltaCPresolution} (left) shows for different $\delta_{CP}$ values the neutrino energy spectrum detected with Super-ORCA after 3 years of running with a 450\,kW beam from Protvino (neutrino mode).
The main $\delta_{CP}$ sensitivity comes from $\nu_e$~CC events, 
which show up to $\sim 40$\% variation in event statistics with $\delta_{CP}$.
The number of $\nu_e$~CC events varies between 8260 (for $\delta_{CP}=
\pi/2$) and 11460 (for $\delta_{CP}=3/2\pi$).
%

The same sensitivity calculation procedure as described in Sec.~\ref{sec:sens_calc} is applied.
The same detector response for the Super-ORCA detector is assumed,
which has not been optimised for a neutrino beam from Protvino,
e.g.\ the known arrival direction of the beam is not exploited (e.g.\ for missing transverse energy).
The atmospheric neutrino flux systematics are replaced with the systematics related to the Protvino neutrino beam flux, as described in \cite{P2O_LoI}.
An equal share between running in neutrino and antineutrino mode was found to be optimal in order to resolve the $\delta_{CP}$-$\theta_{13}$-$\theta_{23}$ degeneracy. 
Note that the neutrino energy spectrum from the Protvino beam has not been optimised for the use of Super-ORCA as far detector. 
A beamline design might improve the $\delta_{CP}$ sensitivity.

Fig.~\ref{fig:atmoANDbeam_fourChi2curves} shows also the $\delta_{CP}$ sensitivity for Super-ORCA using a neutrino beam from Protvino for 10 years (solid lines).
Compared to the measurement with atmospheric neutrinos, the $\delta_{CP}$ sensitivity is significantly larger with up to $12\,\sigma$ between $\delta_{CP}=\pi/2$ and $\delta_{CP}=3/2\pi$.

Fig.~\ref{fig:beam_Nevents_deltaCPresolution} (right) shows the $\delta_{CP}$ resolution that can be achieved after 3 and 10 years of running with a 450\,kW beam.
The 1$\sigma$-resolution is $\sim 10^\circ$ for $\delta_{CP}=0$ and $\delta_{CP}=\pi$, and $\sim 16^\circ$ for $\delta_{CP}=\pi/2$ and $\delta_{CP}=3/2\pi$.
The limiting systematics are the e/$\mu$ energy scale skew, the uncertainty on $\theta_{13}$ and the true value of $\theta_{23}$.
See \cite{P2O_LoI} for further discussions.

\vspace{-0.2cm}
\section{Conclusions}
\label{sec:conclusions}
\vspace{-0.05cm}
The leptonic CP-phase $\delta_{CP}$ can be measured by studying the atmospheric
neutrino oscillation pattern below 2\,GeV with Super-ORCA, 
a $\sim$10 times more-densely instrumented version of KM3NeT/ORCA.
Including systematics,
about 63\% (72\%) of the $\delta_{CP}$ values can be excluded with $\geq 2 \sigma$ 
and a $1\sigma$ resolution on $\delta_{CP}$ of about $38^\circ$ ($23^\circ$) is achieved
for true $\delta_{CP}=0$ and $\delta_{CP}=\pi$ ($\delta_{CP}=\pi/2$ and $\delta_{CP}=3/2\pi$).

The $\delta_{CP}$ sensitivity is significantly improved when using a neutrino beam from 
the Protvino accelerator facility.
A 1$\sigma$-resolution on $\delta_{CP}$ of $\sim 10^\circ$ for $\delta_{CP}=0$ and $\delta_{CP}=\pi$, and $\sim 16^\circ$ for $\delta_{CP}=\pi/2$ and $\delta_{CP}=3/2\pi$ can be achieved after 10 years of running with a 450\,kW beam.

With an energy threshold of few hundreds of MeV, Super-ORCA might also offer interesting possibilities for studying proton decay and Earth neutrino oscillation tomography might become feasible with more than $300$k atmospheric neutrinos per year.

\paragraph*{Acknowledgements}
{\small The authors gratefully acknowledge the support by the DFG (grant EB 569/1-1).}


\end{document}